\documentclass[aps,preprint,pre]{revtex4}
\usepackage{graphics}
\usepackage[dvipdfm]{graphicx}
\usepackage{epsfig}
\usepackage{amsmath}
\begin{document}
\title{Nonlinear Higher-Order Thermo-Hydrodynamics II: Illustrative Examples}
\author{J. Galv\~ao Ramos}
\affiliation{Instituto de F\'{\i}sica ''Gleb Wataghin'', Universidade
Estadual de Campinas - Unicamp, CP 6165, 13083-970, Campinas-SP, Brazil} 
\author{\'Aurea R. Vasconcellos} 
\affiliation{Instituto de F\'{\i}sica ''Gleb Wataghin'', Universidade
Estadual de Campinas - Unicamp, CP 6165, 13083-970, Campinas-SP, Brazil} 
\author{Roberto Luzzi}
\affiliation{Instituto de F\'{\i}sica ''Gleb Wataghin'', Universidade
Estadual de Campinas - Unicamp, CP 6165, 13083-970, Campinas-SP, Brazil} 
\date{\today}
\begin{abstract}
The construction of a generalized (higher-order) nonlinear 
thermo-hydrodynamics, based on a nonequilibrium ensemble formalism has 
been presented in the preceding article. The working of such theory is 
illustrated in the present one. We consider here the case of two ideal 
classical fluids in interaction between them, and the nonlinear 
equations of evolution for the density (a hyperbolic one) and for the 
velocity field describing motion under flow are derived. 
Also, now at the quantum level, it is described the nonlinear transport 
in the fluid of electrons in doped polar semiconductors, and comparison 
with Monte Carlo calculations and with experimental data is done, 
obtaining very good agreement.
\end{abstract}
\maketitle
\section{Introduction}
In the preceding article (hereafter called as I, and when equations 
present in it are indicated in what follows, we write the number 
preceded by I) \cite{a1}, it was described the construction of a 
generalized (higher-order) and nonlinear thermo-hydrodynamics, based on 
mechano-statistical foundations. In that way it was established a 
kind of unification of the microscopic dynamics and a mesoscopic 
hydrodynamics.

In this follow-up article we attempt to illustrate 
the functioning of the theory by applying it to a couple of somewhat 
simplified situations. Both cases are such that we can describe their 
underlying mechanics in terms of an individuals single-particle 
approximation. They are: 1. The case of two ideal classical fluids 
in interaction between them, one being the main system of interest 
driven  away from equilibrium, and the other acting as a thermal 
reservoir; the associated nonlinear generalized hydrodynamics, however 
restricted to a study of a first-order one, is derived in section II. 
2. A similar case, but now composed of two quantum gases, namely, 
conduction band electrons in a doped polar semiconductor under the action 
of an eletric field, and the phonon system acting as a thermal bath, 
is dealt with in section IV. Moreover, the study of section II is 
complemented in section III with the analysis of diffusion-advection 
motion governed by a nonlinear hydrodynamic equation.
\section{Two Classical Fluids in Interaction}
Let us consider two ideal classical fluids with interaction between them, 
with the Hamiltonian given by 
\begin{eqnarray}
H = \sum_{j=1}^{N} \frac{p_{j}^{2}}{2 \, m} + 
\sum_{\mu=1}^{N_{R}} \frac{P_{\mu}^{2}}{2 \, M} + 
\sum_{j,\mu} {\mathcal V}(|{\mathbf r}_{j} - {\mathbf R}_{\mu}|) + 
V_{ext} \, ,
\end{eqnarray}
$m$ being the mass of the particles in the system of interest, $M$ that in 
the second system taking as a thermal bath at temperature $T_{0}$, and the 
interaction being of the type of central forces. $V_{ext}$ stands for the 
interaction energy with an external source, and it is to be understood that 
the bath is constantly in equilibrium with an external ideal reservoir. We 
analize the {\it hydrodynamics of this system in a first order approximation}, meaning to 
take as basic variables of the system of interest the kinetic energy, 
the density, and the first flux (current) of the latter, with the 
associated dynamical quantities being
\begin{eqnarray}
\hat{h}({\mathbf r}) = \sum_{j} \frac{p_{j}^{2}}{2 \, m} \, 
\delta({\mathbf r} - {\mathbf r}_{j}) \, ,
\end{eqnarray}
\begin{eqnarray}
\hat{n}({\mathbf r}) = \sum_{j} \delta({\mathbf r} - {\mathbf r}_{j}) \, ,
\end{eqnarray}
\begin{eqnarray}
\hat{{\mathbf I}}_{n}({\mathbf r}) = \sum_{j} \frac{{\mathbf p}_{j}}{m} \, 
\delta({\mathbf r} - {\mathbf r}_{j}) \, ,
\end{eqnarray}
with the understanding that all three are defined in $\Gamma$-phase space. 
The thermal bath is assumed to remain constantly in equilibrium with a 
reservoir at temperature $T_{0}$, and the system subjected to external 
forces driving it out of equilibrium.

According to I (the previous article) the auxiliary (''instantaneously 
frozen'') nonequilibrium statistical operator is given by [cf. Eq.(I.6)] 
\begin{eqnarray}
\bar{{\mathcal R}}(t,0) = \bar{\rho}(t,0) \times \rho_{R} \, , 
\end{eqnarray}
where $\rho_{R}$ is the canonical distribution at temperature $T_{0}$ of 
the thermal bath, and 
\begin{eqnarray}
\bar{\rho}(t,0) = \exp \left\{ - \phi(t) - \int d^{3}r \left[ 
F_{h}({\mathbf r},t) \, \hat{h}({\mathbf r}) + 
A({\mathbf r},t) \, \hat{n}({\mathbf r}) + 
{\mathbf F}_{n}({\mathbf r},t) \cdot \hat{{\mathbf I}}_{n}({\mathbf r}) 
\right] \right\} \, ,
\end{eqnarray}
where, we recall, $\phi(t)$ ensures the normalization and we have 
introduced the set of {\it nonequilibrium thermodynamic variables} (Lagrange 
multipliers in the variational derivation of the formalism), namely
\begin{eqnarray}
\left\{ F_{h}({\mathbf r},t), A({\mathbf r},t), {\mathbf F}_{n}({\mathbf r},t) 
\right\} \, ,
\end{eqnarray}
and we also recall that the nonequilibrium statistical operator is given, in 
terms of the auxiliary one of Eq.(5), by Eq.(I.4). The corresponding set of 
{\it basic hydrodynamic macrovariables} consists of 
\begin{eqnarray}
h({\mathbf r},t) = \int d\Gamma \, \hat{h}({\mathbf r}) \, \bar{\rho}(t,0) \, , 
\end{eqnarray}
\begin{eqnarray}
n({\mathbf r},t) = \int d\Gamma \, \hat{n}({\mathbf r}) \, \bar{\rho}(t,0) \, , 
\end{eqnarray}
\begin{eqnarray}
{\mathbf I}_{n}({\mathbf r},t) = 
\int d\Gamma \, \hat{{\mathbf I}}_{n}({\mathbf r},t)
\, \bar{\rho}(t,0) \, , 
\end{eqnarray}
where the integration is over the phase space of the system [cf. Eq.(I.31)]. 
These equations, Eqs.(8) to (10), are the {\it nonequilibrium thermodynamic equations 
of state}, meaning that they relate the basic macrovariables on the left, with 
the nonequilibrium thermodynamic variables of Eq.(7) present on the right-hand 
side in the distribution $\bar{\rho}$ of Eq.(6). 

Equations (8) to (10) take the form
\begin{eqnarray}
h({\mathbf r},t) = \int d^{3}p \, \frac{p^{2}}{2 \, m} \, 
f({\mathbf r},{\mathbf p};t) \, , 
\end{eqnarray}
\begin{eqnarray}
n({\mathbf r},t) = \int d^{3}p \, f({\mathbf r},{\mathbf p};t) \, , 
\end{eqnarray}
\begin{eqnarray}
{\mathbf I}_{n}({\mathbf r},t) = \int d^{3}p \, \frac{{\mathbf p}}{m}  
f({\mathbf r},{\mathbf p};t) \, , 
\end{eqnarray}
where
\begin{eqnarray}
f({\mathbf r},{\mathbf p};t) = \frac{1}{{\mathcal Z}(t)} \exp \left\{ 
A({\mathbf r},t) - F_{h}({\mathbf r},t) \, \frac{p^{2}}{2 \, m} - 
{\mathbf F}_{n}({\mathbf r},t) \cdot \frac{{\mathbf p}}{m} \right\} \, ,
\end{eqnarray}
is a Boltzmann-style one-particle distribution function, and 
\begin{eqnarray}
{\mathcal Z}(t) = \int d^{3}r  \int d^{3}p \, \exp \left\{ 
A({\mathbf r},t) - F_{h}({\mathbf r},t) \, \frac{p^{2}}{2 \, m} + 
{\mathbf F}_{n}({\mathbf r},t)  \cdot \frac{{\mathbf p}}{m} \right\} \, .
\end{eqnarray}
Performing the calculations we find that 
\begin{eqnarray}
n({\mathbf r},t) = \frac{N}{{\mathcal Z}(t)} 
\left( \frac{2 \, \pi \, m}{F_{h}({\mathbf r},t)} \right)^{3/2} \, \exp \left\{
A({\mathbf r},t) + F_{h}({\mathbf r},t) \, \frac{1}{2} m \, 
v^{2}({\mathbf r},t) \right\} \, , 
\end{eqnarray}
where we have introduced the barycentric velocity ${\mathbf v}({\mathbf r},t)$ 
[cf. Eq.(20) below], through the definition (see also Eq.(20) below) 
\begin{eqnarray}
{\mathbf F}_{n}({\mathbf r},t) = F_{h}({\mathbf r},t) \, 
m \, {\mathbf v}({\mathbf r},t) \, .
\end{eqnarray}

Moreover, we do find that 
\begin{eqnarray}
h({\mathbf r},t) = u({\mathbf r},t) + 
n({\mathbf r},t) \frac{1}{2} m \, v^{2}({\mathbf r},t) \, ,
\end{eqnarray}
for the energy, where $u$ is the internal energy 
\begin{eqnarray}
u({\mathbf r},t) = \frac{3}{2} \frac{n({\mathbf r},t)}{F_{h}({\mathbf r},t)} = 
\frac{3}{2} n({\mathbf r},t) \, k_{B} \, T^{\ast}({\mathbf r},t) \, , 
\end{eqnarray}
introducing, via the definition $F_{h}^{- 1} = k_{B} \, T^{\ast}$, 
the field of quasitemperature \cite{a2,a3,a4} or nonequilibrium kinetic 
temperature \cite{a5} $T^{\ast}({\mathbf r},t)$, with the last term on the 
right of Eq.(18) being the drift-kinetic energy, and 
\begin{eqnarray}
{\mathbf I}_{n}({\mathbf r},t) = n({\mathbf r},t) \, 
{\mathbf v}({\mathbf r},t) .\,  
\end{eqnarray}

It can be noticed that if $n$ is the concentration of particles, it needs be 
satisfied the constraint that 
\begin{eqnarray}
n = \int d^{3}r \, n({\mathbf r},t) \, ,
\end{eqnarray}
with $n({\mathbf r},t)$ given by Eq.(12).

Let now go over the generalized hydrodynamic equations, namely those of 
Eqs.(I.35), (I.36) and (I.38) of I (previous article). We do have that 
\begin{eqnarray}
\frac{\partial \hfill}{\partial t} h({\mathbf r},t) + 
\nabla \cdot {\mathbf I}_{h}({\mathbf r},t) = J_{h}({\mathbf r},t) \, ,
\end{eqnarray}
where ${\mathbf I}_{h}$ is the flux of energy and $J_{h}$ is the collision 
integral; 
\begin{eqnarray}
\frac{\partial \hfill}{\partial t} n({\mathbf r},t) + 
\nabla \cdot {\mathbf I}_{n}({\mathbf r},t) = 0 \, ,
\end{eqnarray}
which is the conservation equation for the density; 
\begin{eqnarray}
\frac{\partial \hfill}{\partial t} {\mathbf I}_{n}({\mathbf r},t) + 
\nabla \cdot I_{n}^{[2]}({\mathbf r},t) = {\mathbf J}_{n}({\mathbf r},t) + 
{\mathbf F}_{ext}({\mathbf r},t) \, ,
\end{eqnarray}
where $I_{n}^{[2]}({\mathbf r},t)$ is the flux of flux (second-order flux) of 
particles, namely
\begin{eqnarray}
I_{n}^{[2]}({\mathbf r},t) = \int d\Gamma \, \sum_{j} 
\left[ \frac{{\mathbf p}_{j}}{m} \frac{{\mathbf p}_{j}}{m} \right] 
\delta( {\mathbf r} - {\mathbf r}_{j} ) \, \bar{\rho}(t,0) \, ,
\end{eqnarray}
with $[ {\mathbf p}_{j} \, {\mathbf p}_{j} ]$ standing for tensorial product 
of vectors, rendering a second-rank tensor, ${\mathbf F}_{ext}({\mathbf r},t)$ 
stands for external forces [arising out of $V_{ext}$ in Eq.(1)] applied on 
the system driven it out of equilibrium, and ${\mathbf J}_{n}$ is the 
collision integral accounting for relaxation processes (towards the thermal 
bath): See Appendix A. The latter is composed of three contributions: two of 
them, one introducing a term proportional to the gradient of concentration 
(and responsible for a perturbational modification of the diffusion effect; 
cf. Ref.[6]) and another introducing nonlocal effects (correlations in space), 
are neglected and we conserve the main contribution given by 
\begin{eqnarray}
{\mathbf J}_{n}({\mathbf r},t) = 
- \left[ \Theta_{n}^{[2]}({\mathbf r},t) \right]^{- 1} \, 
{\mathbf I}_{n}({\mathbf r},t) \, ,
\end{eqnarray}
introducing the inverse of a second-rank tensor, $\Theta_{n}^{[2]}$ having 
dimension of time and playing the role of a tensorial-in-character 
Maxwell-relaxation time \cite{a4,a7,a8}, its explicit expression given in 
Eq.(A3) of Appendix A. 

Finally, to close the system of Eqs.(22) to (24) coupled to the set of 
Eqs.(16), (18) and (20) - the equations of state - we need to express the 
second-order flux of Eq.(25) in terms of the basic variables, and we obtain 
that
\begin{eqnarray}
m \, I_{n}^{[2]}({\mathbf r},t) = n({\mathbf r},t) \left[ 
k_{B} \, T^{\ast}({\mathbf r},t) \, 1^{[2]} + 
m \, n({\mathbf r},t) \, \left[ {\mathbf v}({\mathbf r},t) 
{\mathbf v}({\mathbf r},t) \right] \right] \, .
\end{eqnarray}

This second-order flux is related to the field of pressure tensor by 
\begin{eqnarray}
P^{[2]}({\mathbf r},t) = m \, I_{n}^{[2]}({\mathbf r},t) - 
n({\mathbf r},t) \left[ {\mathbf v}({\mathbf r},t) {\mathbf v}({\mathbf r},t) 
\right] = n({\mathbf r},t) \, k_{B} \, T^{\ast}({\mathbf r},t) \, 1^{[2]} = 
\frac{2}{3} u({\mathbf r},t) \, 1^{[2]} \, ,
\end{eqnarray}
after using Eq.(19), recovering the local and instantaneous usual form of the 
hydrostatic pressure.

This first-order hydrodynamics of this simple modeled system sets clearly in 
evidence the {\it nonlinearity} referred to in I, namely, one coming out of 
the expression for the second flux, cf. Eq.(27), and other the higher nonlinear 
expression for the collision integrals. We also stress that in Eq.(26) 
$\Theta_{n}^{[2]}({\mathbf r},t)$ is also highly nonlinear in the 
basic thermodynamic variables.

In that way it has been explicitly shown that nonlinearities are always present 
in the generalized hydrodynamic equations, and we further illustrate the matter 
first, on the basis of the results above, deriving nonlinear equations of 
evolution for the density and the velocity in the fluid under flow, and after 
that a study of electric current in a system of mobile electrons in a doped 
polar semiconductor, dealt with at the quantum mechanical level.
\section{Nonlinear Equations of Evolution}
Let us consider the equations of evolution for the basic variables of Eqs.(8) 
to (10), given by Eqs.(22), (23) and (24). Deriving in time Eq.(23) and using 
Eq.(24) it follows that
\begin{eqnarray}
\frac{\partial^{2} \hfill}{\partial t^{2}} n({\mathbf r},t) =  
- \nabla \cdot \frac{\partial \hfill}{\partial t} 
{\mathbf I}_{n}({\mathbf r},t) = 
\nabla \cdot \nabla \cdot I_{n}^{[2]}({\mathbf r},t) + 
\nabla \cdot \frac{{\mathbf I}_{n}({\mathbf r},t)}{\Theta_{n}({\mathbf r},t)} - 
\nabla \cdot {\mathbf F}_{ext}({\mathbf r},t) \, ,
\end{eqnarray}
where for simplicity we have taken the tensorial Maxwell characteristic time 
of Eq.(26) as a scalar (see Appendix A).

Equation (29) can be rewritten as 
\begin{eqnarray}
&&\Theta_{n}({\mathbf r},t) \, 
\frac{\partial^{2} \hfill}{\partial t^{2}} n({\mathbf r},t) +   
\frac{\partial \hfill}{\partial t} n({\mathbf r},t) + 
\nabla \cdot \left[ n({\mathbf r},t) \, {\mathbf v}_{a}({\mathbf r},t) 
\right] = \nonumber \\
&&\Theta_{n}({\mathbf r},t) \, \nabla \cdot \nabla \cdot 
I_{n}^{[2]}({\mathbf r},t) - 
\Theta_{n}({\mathbf r},t) \, {\mathbf I}_{n}({\mathbf r},t) \cdot 
\nabla \Theta_{n}^{- 1}({\mathbf r},t) \, ,
\end{eqnarray}
after multiplying by $\Theta$ and the definition 
\begin{eqnarray}
\Theta_{n}({\mathbf r},t) \, \nabla \cdot 
{\mathbf F}_{ext}({\mathbf r},t) \equiv 
\nabla \cdot \left[ n({\mathbf r},t) \, {\mathbf v}_{a}({\mathbf r},t) 
\right] \, ,
\end{eqnarray}
introducing ${\mathbf v}_{a}$ which we call velocity responsible for creating a 
driven flow in the system. 

Second, in order to close Eq.(30) we need to express the second-order flux in 
terms of the basic variables, $n$ and ${\mathbf I}_{n}$, which after some 
calculus takes the form 
\begin{eqnarray}
I_{n}^{[2]}({\mathbf r},t) = n({\mathbf r},t) \, k_{B} \, 
T^{\ast}({\mathbf r},t) \, 1^{[2]} + 
n({\mathbf r},t) \left[ {\mathbf v}({\mathbf r},t) {\mathbf v}({\mathbf r},t) 
\right] \, .
\end{eqnarray}

Using this Eq.(32) in Eq.(30) and, for further simplifying the presentation, 
let us take $\Theta_{n}({\mathbf r},t)$ as smoothly dependent on ${\mathbf r}$ 
and $t$, i.e. its gradient and time derivative can be neglected, we finally 
obtain that
\begin{eqnarray}
&&\Theta_{n}({\mathbf r},t) \, 
\frac{\partial^{2} \hfill}{\partial t^{2}} n({\mathbf r},t) +   
\frac{\partial \hfill}{\partial t} n({\mathbf r},t) - 
{\mathcal D}_{n}({\mathbf r},t) \nabla^{2} n({\mathbf r},t) = \nonumber \\
&&- \nabla \cdot 
\left[ n({\mathbf r},t) \, {\mathbf v}_{a}({\mathbf r},t) \right] + 
\Theta_{n}({\mathbf r},t) \, \nabla \cdot \nabla \cdot \Big( 
n({\mathbf r},t) \, \left[ {\mathbf v}({\mathbf r},t) 
{\mathbf v}({\mathbf r},t) \right] \Big) + G_{n}({\mathbf r},t) \, ,
\end{eqnarray}
where $G_{n}$ contains gradients and time derivatives of 
$\Theta_{n}({\mathbf r},t)$ which, for simplicity we ignore in what follows, 
\begin{eqnarray}
{\mathcal D}_{n}({\mathbf r},t) = \frac{1}{3} v^{2}({\mathbf r},t) \, 
\Theta_{n}({\mathbf r},t) \, ,
\end{eqnarray}
we have used that ${\mathbf I}_{n}({\mathbf r},t) = n({\mathbf r},t) \, 
{\mathbf v}({\mathbf r},t)$, and the second contribution on the 
right-hand side - arising out of the contribution of the so-called convective 
pressure - can be written as
\begin{eqnarray}
\nabla \cdot \nabla \cdot \Big( n({\mathbf r},t) \, 
\left[ {\mathbf v}({\mathbf r},t) {\mathbf v}({\mathbf r},t) 
\right] \Big) &=& \Big( {\mathbf v}({\mathbf r},t) \cdot \nabla \Big)^{2} \, 
n({\mathbf r},t) + 2 \Big( {\mathbf v}({\mathbf r},t) \cdot \nabla 
n({\mathbf r},t) \Big) \nabla \cdot {\mathbf v}({\mathbf r},t) \nonumber \\
&&+ \, 
n({\mathbf r},t) \Big( \nabla \cdot {\mathbf v}({\mathbf r},t) \Big)^{2} + 
2 \, n({\mathbf r},t) \Big( {\mathbf v}({\mathbf r},t) \cdot \nabla \Big) 
\Big( \nabla \cdot {\mathbf v}({\mathbf r},t) \Big) \nonumber \\
&&+ \,
n({\mathbf r},t) \tilde{\Lambda}^{[2]}({\mathbf r},t) \otimes 
\tilde{\Lambda}^{[2]}({\mathbf r},t) \nonumber \\
&&+ \,
\Big[ {\mathbf v}({\mathbf r},t) : \nabla n({\mathbf r},t) \Big] \otimes 
\tilde{\Lambda}^{[2]}({\mathbf r},t) \, ,
\end{eqnarray}
where $\tilde{\Lambda}^{[2]}$ is the velocity-gradient tensor, 
$\tilde{\Lambda}^{[2]} = [ \nabla {\mathbf v}({\mathbf r},t) ]$. 

Hence, the equation of evolution for the density, Eq.(3), is dependent on 
the velocity field and then it needs be coupled to the equation of evolution 
for the velocity, which is
\begin{eqnarray}
n({\mathbf r},t) \left[ \frac{\partial \hfill}{\partial t} + 
{\mathbf v}({\mathbf r},t) \cdot \nabla \right] \, {\mathbf v}({\mathbf r},t) 
= - \frac{1}{m} \nabla \cdot P^{[2]}({\mathbf r},t) - 
n({\mathbf r},t) \frac{{\mathbf v}({\mathbf r},t) -  
{\mathbf v}_{a}({\mathbf r},t)}{\tau_{n}({\mathbf r},t)} \, ,
\end{eqnarray}
where the pressure tensor field is given by $P^{[2]}({\mathbf r},t) = 
n({\mathbf r},t) \, k_{B} \, T^{\ast}({\mathbf r},t) \, 1^{[2]}$.

Equation (33) is a nonlinear Maxwell-Cattaneo-like equation with sources. 
Moreover, in the strong condition of neglecting the right-hand side and the 
second derivative in time implying in a movement in conditions such that 
$\omega \, \Theta_{n} \ll 1$, we recover the standard Fick's diffusion equation 
\begin{eqnarray}
\frac{\partial \hfill}{\partial t} n({\mathbf r},t) - 
{\mathcal D}_{n}({\mathbf r},t) \, \nabla^{2} n({\mathbf r},t) = 0 \, .
\end{eqnarray}

Next let us introduce the effects of advection in the diffusive regime, i.e. 
the density follows the diffusion-advection equation of evolution, as given 
by Eq.(33), in the condition $\omega \, \Theta_{n} \ll 1$ (thus neglecting 
the second derivative in time), and for simplicity we illustrate the matter 
in the case of motion restricted to proceed in $x$-direction; hence we do 
have that 
\begin{eqnarray}
\frac{\partial \hfill}{\partial t} n(x,t) - 
{\mathcal D}_{n}(x,t) \,  \frac{\partial^{2} \hfill}{\partial x^{2}}n(x,t) = 
- \frac{\partial \hfill}{\partial x} \Big[ n(x,t) \, v_{a}(x,t) \Big] +
\Theta_{n}(x,t) \,  \frac{\partial^{2} \hfill}{\partial x^{2}} 
\Big[ n(x,t) \, v^{2}(x,t) \Big] \, .
\end{eqnarray}
Moreover, the diffusion coefficient has the standard expression of kinetic 
theory (in one dimension), ${\mathcal D}_{n} = v_{th}^{2} \, \Theta_{n}$, 
where $v_{th}$ is the local and instantaneous thermal velocity 
$m \, v_{th}^{2}(x,t) = k_{B} \, T^{\ast}(x,t)$. 

Equation (38) can be reorganized in the following way: 
\begin{eqnarray}
\frac{\partial \hfill}{\partial t} n(x,t) &-& 
{\mathcal D}_{n}(x,t) \,  \frac{\partial^{2} \hfill}{\partial x^{2}}n(x,t) = 
- \frac{\partial \hfill}{\partial x} \Big[ n(x,t) \, v_{a}(x,t) \Big] +
2 \, \Theta_{n}(x,t) \, n(x,t) \, v(x,t) 
\frac{\partial^{2} v^{2}(x,t) \hfill}{\partial x^{2}} \nonumber \\
&&+ \, 2 \, \Theta_{n}(x,t) \, n(x,t) \left[ \frac{\partial v(x,t) 
\hfill}{\partial x}  \right]^{2} + 
2 \, \Theta_{n}(x,t) \, v(x,t) \frac{\partial \hfill}{\partial x} n(x,t) 
\frac{\partial \hfill}{\partial x} v(x,t) \, ,
\end{eqnarray}
where
\begin{eqnarray}
{\mathcal D}_{n}(x,t) = \Theta_{n}(x,t) \, \left[ v_{th}^{2}(x,t) + v^{2}(x,t) 
\right] \, , 
\end{eqnarray}
that is, a diffusion coefficient whose standard kinetic theory expression is 
modified adding to the contributions of the thermal velocity the one of the 
drift velocity. This contribution together with the last three on the 
right-hand side of Eq.(39) have their origin in the second divergence of the 
convection pressure, and it can be noticed that all are nonlinear (quadratic) 
in $v(x,t)$. The first term on the right of Eq.(39), we recall, is related to 
the divergence of the driving force producing the advective motion 
[cf. Eq.(31)]. 

Moreover, as already noticed, Eq.(39) needs be solved in conjunction with the 
equation of evolution for the velocity, in this case of motion under flow 
given by
\begin{eqnarray}
&&n(x,t) \frac{\partial \hfill}{\partial t} v(x,t) + 
n(x,t) \, v(x,t) \frac{\partial \hfill}{\partial x} v(x,t) = \nonumber \\
&&- \, \frac{k_{B}}{m} \frac{\partial \hfill}{\partial x} 
T^{\ast}(x,t) \, n(x,t) - 
n(x,t) \, \Theta_{n}^{- 1}(x,t) \, \left[ v(x,t) - v_{a}(x,t) \right] \, , 
\end{eqnarray}
where we used that $p(x,t) = n(x,t) \, k_{B} \, T^{\ast}(x,t)$. 
\section{Nonlinear Quantum Transport in Semiconductors}
We consider the case of polar semiconductors described by a 
two-inverted-parabolic bands model (in the effective mass approximation and 
conduction band secondary valleys are ignored), where a concentration $n_{e}$ 
of mobile electrons in the conduction band has been created by doping. 
A constant eletric field of intensity $E_{0}$ in, say, $x$-direction 
is applied, which accelerates the electrons (''hot'' electrons) while there 
follows a transferring of their energy and momentum (in excess of equilibrium) 
to the phonon field. The sample is in contact with a thermal reservoir at 
temperature $T_{0}$, with the phonons being warmed up in scattering events 
involving Fr\"ohlich, deformation potential, and piezoeletric interactions 
with the ''hot'' electrons \cite{a9,a10} (see also Ch. 6 in the book of 
Ref.[11]). Scattering by impurities is neglected in comparison with the one 
due to lattice vibrations. Moreover, in these polar semiconductors of the 
different types of electron-phonon interaction we keep only the predominat 
Fr\"ohlich (polar) interaction. 

The electric field ${\mathbf E}$ creates a (uniform in space) current, 
that is, a flux of charged particles, and then we have a kind of hydrodynamic 
motion of the like of the one analysed in the previous sections, except 
that, as noticed, the hydrodynamic variables are uniform in space (independent 
of position ${\mathbf r}$), and the thermal bath is played by the phonons. 
We resort to a first-order hydrodynamics, as in section II, namely, we 
consider the carriers' energy, the density of particles and the flux of
particles. The latter multiplied by the effective mass of the conduction band 
electrons is the linear momentum, and while the density of particles remains 
constant in time, as noticed the energy and the momentum change in time. 
The Hamiltonian is composed of the energies of the electrons in the conduction 
band (treated as usual in the random-phase approximation thus including Coulomb 
interaction in a mean-field approximation) and that of the polar (longitudinal 
optical) phonons, we call $\hat{H}_{0}$ this part, plus the Fr\"ohlich 
interaction of electrons and polar phonons and the interaction of the electrons 
with the electric field, we call $\hat{H}^{\prime}$ this part. Moreover, 
differently in other aspect with the case of section II we need here 
to use a quantum approach.

Taking as basic dynamical variables $\hat{H}_{e}$, $\hat{H}_{LO}$, 
$\hat{N}$, $\hat{{\mathbf P}}$ (energy of electrons $\hat{H}_{e}$ and 
polar phonons $\hat{H}_{LO}$, number of electrons, and linear momentum 
of the electrons) the auxiliary (''instantaneously frozen'') statistical 
operator is then given by
\begin{eqnarray}
\bar{\rho}(t,0) = \exp \left\{ - \phi(t) - \beta(t) \left[ 
\hat{H}_{e} - \mu(t) \, \hat{N} - {\mathbf v}(t) \cdot \hat{{\mathbf P}} 
\right] - \beta_{LO}(t) \, \hat{H}_{LO} \right\} \, \rho_{R} \, ,
\end{eqnarray}
where, we recall, $\phi(t)$ ensures the normalization condition, and we 
have introduced the nonequilibrium thermodynamic variables
\begin{eqnarray}
\Big\{ \beta(t) \equiv 1 / k_{B} T^{\ast}(t), - \beta(t) \, \mu(t), 
- \beta(t) \, {\mathbf v}(t), \beta_{LO}(t) \equiv 
1/ k_{B} T_{LO}^{\ast}(t) \Big\} \, ,
\end{eqnarray}
interpreted as nonequilibrium quasitemperature, $T^{\ast}$, quasi-chemical 
potential, $\mu$, and drift velocity, ${\mathbf v}$, of the electrons, and 
quasitemperature of the polar phonons, $T_{LO}^{\ast}$. Finally, $\rho_{R}$ is 
the statistical operator of the reservoir at temperature $T_{0}$. 

We call the basic macrovariables 
\begin{eqnarray}
\Big\{ E_{e}(t), N, {\mathbf P}(t), E_{LO}(t) \Big\} \, ,
\end{eqnarray}
where $N$ is constant in time, and we recall, the density is indicated by $n$, 
and the nonequilibrium thermodynamic equations of state are, after direct 
calculation, given by
\begin{eqnarray}
\frac{1}{V} E_{e}(t) = \frac{3}{2} n \, k_{B} \, T^{\ast}(t) + 
\frac{1}{2} n \, m^{\ast} \, v^{2}(t) \, ,
\end{eqnarray}
\begin{eqnarray}
\frac{1}{V} {\mathbf P}(t) = n \, m^{\ast} \, {\mathbf v}(t) \, ,
\end{eqnarray}
\begin{eqnarray}
\frac{1}{V} E_{LO}(t) = \frac{1}{V_{cell}} \hbar \, \omega_{0} \Big[ 
\exp \left\{ \beta_{LO}(t) \, \hbar \, \omega_{0} \right\} - 1 \Big]^{- 1} \, ,
\end{eqnarray}
where, 1. In the conditions of concentration and values of quasitemperature 
involved in the applications, the nonequilibrium time-dependent 
Fermi-Dirac-like distribution of the internally thermalized electrons in band 
states in the effective mass approximation can be approximated by a 
nonequilibrium time-dependent Boltzmann-Maxwell one \cite{a9,a11}, and then 
there follows the first term on the right of Eq.(45), namely, equipartition 
at each time $t$ in terms of the quasitemperature $T^{\ast}(t)$ of the 
''hot'' electrons of effective mass $m^{\ast}$, and the other term is 
evidently the kinetic energy of drift (current). 2. The polar phonons have 
been treated in a dispersionless approximation (Einstein model), with unique 
frequency $\omega_{0}$, $V_{cell}$ is the volume of the unit cell in the 
crystal, and $V$ is the volume of the sample.

On the other hand the equations of evolution (see Appendix B) are
\begin{eqnarray}
\frac{d \hfill}{dt} E_{e}(t) = \frac{e}{m^{\ast}} {\mathbf E} \cdot 
{\mathbf P}(t) - \left| J_{E}(t) \right| \, , 
\end{eqnarray}
for the energy of the electrons, 
\begin{eqnarray}
\frac{d \hfill}{dt} {\mathbf P}(t) = e \, n \, {\mathbf E} - 
\left| {\mathbf J}_{P}(t) \right| \, , 
\end{eqnarray}
for the electron linear momentum in the direction of the applied field, 
\begin{eqnarray}
\frac{d \hfill}{dt} E_{LO}(t)= \left| J_{P}(t) \right| - 
\left| J_{an}(t) \right| \, ,
\end{eqnarray}
for the energy of the polar phonons; volume $V$ has been taken as $1$.

In Eq.(48) the first term on the right accounts for the pumping of energy 
on the carrier system because of the presence of the electric field, while 
the second represents the rate of excess energy transferred to the polar 
phonons. In Eq.(49) the first term on the right is the force produced by 
the presence of the electric field, and the second the rate of momentum 
transferred to the lattice. In Eq.(50) the first term on the right is the 
gain of energy pumped on the phonons by the nonequilibrated (''hot'') 
carriers, with the second being the transfer - via anharmonic processes - 
of such energy to the acoustic phonons acting as a thermal bath (finally, 
from the latter there follows a transfer to the thermal reservoir). As 
noticed, their expressions are given in the Appendix B with the calculations 
being performed in the Markovian \cite{a24,a25,a26}. We write down here 
the two collision integrals of Eqs.(48) and (49) to illustrate the 
{\it highly nonlinear dependence} of them on the nonequilibrium thermodynamic 
variables, $\beta(t)$ and ${\mathbf v}(t)$, which, on the other hand, are 
related to the basic variables through the also {\it nonlinear equations of state}, 
viz. Eqs.(45) and (46): 
\begin{eqnarray}
\left| J_{E}(t) \right| = {\mathcal A}_{E} \, y^{3/2} \, e^{- x} \, n \, 
\Big[ \nu_{0} \, A_{1} - \left( \nu_{0} + 1 \right) A_{2} \Big] \, ,
\end{eqnarray}
\begin{eqnarray}
\left| J_{P}(t) \right| = {\mathcal A}_{P} \, y^{3/2} \, e^{- x} \, n \, 
\Big[ \nu_{0} \, A_{3} - \left( \nu_{0} + 1 \right) A_{4} \Big] \, ,
\end{eqnarray}
where
\begin{eqnarray}
y(t) = \beta(t) \, \hbar \, \omega_{0} 
\: \: \: \: \: \: ; \: \: \: \: \: \: 
x(t) = \frac{1}{2} m^{\ast} \, v^{2}(t) / k_{B} \, T^{\ast}(t) \, , 
\end{eqnarray}
\begin{eqnarray}
{\mathcal A}_{E} = \left( \frac{2 \, \hbar \, \omega_{0}}{\pi \, m^{\ast}} 
\right)^{1/2} \, e \, {\mathcal E}_{0} 
\: \: \: \: \: \: ; \: \: \: \: \: \:
{\mathcal A}_{P} = \frac{e \, {\mathcal E}_{0}}{2 \sqrt{\pi}} \, , 
\end{eqnarray}
\begin{eqnarray}
\nu_{0} = \Big[ \exp \left\{ \beta_{LO}(t) \, \hbar \, \omega_{0} \right\} - 1 
\Big]^{- 1} \, , 
\end{eqnarray}
\begin{eqnarray}
A_{1}(t) = \frac{e^{y/2}}{y} K_{0}(y/2) + 
\sum_{n=1}^{\infty} \frac{(4 \, x \, y)^{n}}{(2n+1)!} (- 1)^{n} 
\frac{d^{n} \hfill}{dy^{n}} \left[ \frac{e^{y/2}}{y} \, K_{0}(y/2) \right] \, , 
\end{eqnarray}
\begin{eqnarray}
A_{2}(t) = e^{- y} \, \frac{e^{y/2}}{y} K_{0}(y/2) + 
\sum_{n=1}^{\infty} \frac{(4 \, x \, y)^{n}}{(2n+1)!} (- 1)^{n} 
\frac{d^{n} \hfill}{dy^{n}} \left[ \frac{e^{y/2}}{y} \, K_{0}(y/2) \right] \, , 
\end{eqnarray}
\begin{eqnarray}
A_{3}(t) &=& \frac{1}{3} (4 \, x \, y)^{1/2} \frac{e^{y/2}}{y} 
\Big[ K_{0}(y/2) - K_{1}(y/2) \Big] \nonumber \\
&&+ 
\sum_{n=2}^{\infty} \frac{2n}{(2n+1)!} \, (4 \, x \, y)^{\frac{2n-1}{2}}  
(- 1)^{n-1} \, \frac{d^{n-1} \hfill}{dy^{n-1}} \left[ 
\frac{e^{y/2}}{y} \, \Big( K_{0}(y/2) - K_{1}(y/2) \Big) \right] \, , 
\end{eqnarray}
\begin{eqnarray}
A_{4}(t) &=& \frac{1}{3} e^{- y} \, (4 \, x \, y)^{1/2} \frac{e^{y/2}}{y} 
\Big[ K_{0}(y/2) + K_{1}(y/2) \Big] \nonumber \\
&&+ 
\sum_{n=2}^{\infty} \frac{2n}{(2n+1)!} \, (4 \, x \, y)^{\frac{2n-1}{2}}  
(- 1)^{n-1} \, \frac{d^{n-1} \hfill}{dy^{n-1}} \left[ 
\frac{e^{y/2}}{y} \, \Big( K_{0}(y/2) + K_{1}(y/2) \Big) \right] \, . 
\end{eqnarray}
In these expressions - and note that we have omitted on the right-hand side 
to explicitly indicate the dependence on time of the different quantities 
involved - $K_{n}$ stands for Bessel function of order $n$ \cite{a12} 
and ${\mathcal E}_{0}$ is the Fr\"ohlich field intensity in the polar 
interaction \cite{a13}. 

It is worth noticing that in quantities $A_{3}$ and $A_{4}$ [cf. Eqs.(58) 
and (59)] it can be put in evidence the quantity $x^{1/2}$ which is 
proportional to $v$, and therefore we can formally write Eq.(49) as 
\begin{eqnarray}
\frac{d \hfill}{dt}v(t) = \frac{e}{m^{\ast}} E_{0} - 
\frac{v(t)}{\tau_{_{P}}(t)} \, ,
\end{eqnarray}
after using Eq.(46), being an equation of the Newton-Langevin type but where 
$\tau_{_{P}}(t)$ plays the role of a relaxation time of the velocity (momentum) 
which is depending on time through its dependence on the nonequilibrium 
thermodynamic state of the system, i.e. depends on a highly nonlinear way on 
$\beta(t)$, $\beta_{LO}(t)$, and the velocity $v(t)$. Equation (60) can be 
alternatively written in the form of the integral equation
\begin{eqnarray}
v(t) = \frac{e}{m^{\ast}} E_{0} \, \tau_{_{c}}(t) \, ,
\end{eqnarray}
where
\begin{eqnarray}
\tau_{_{c}}(t) = \exp \left\{ - \psi(t) \right\} \, \int_{0}^{\infty} dt' \, 
\exp \left\{ \psi(t') \right\} \, ,
\end{eqnarray}
with 
\begin{eqnarray}
\psi(t) = \int_{0}^{t} dt' \, \tau_{_{P}}^{- 1}(t') \, ,
\end{eqnarray}
after taking into account the initial condition $v(0) = 0$. Moreover, once 
the current density is given by 
\begin{eqnarray}
I(t) = e \, n \, v(t) \, ,
\end{eqnarray}
using Eq.(61) we can write 
\begin{eqnarray}
I(t) = \sigma(t) \, E_{0} \, ,
\end{eqnarray}
introducing a time-dependent (on the evolution of the nonequilibrium 
thermodynamic state of the system) Drude-type conductivity 
\begin{eqnarray}
\sigma(t) = \frac{n \, e^{2}}{m^{\ast}} \, \tau_{_{c}}(t) \, .
\end{eqnarray}

Next we illustrate numerically the matter considering the polar semiconductor 
GaAs and the strong polar ones and large gap GaN (of present interest 
for its use in blue diodes and lasers \cite{a14}). Figure 1 shows the increase, 
in the steady state (which follows after a transient of a few picoseconds), of 
the electrons' quasitemperature with the electric field strength in the case 
of doped ($n \simeq 10^{16}cm^{- 3}$) GaAs at a reservoir temperature 
$T_{0} = 300K$. The dots are the result of a Nonequilibrium Molecular 
Dynamics (Monte Carlo-style) simulation (from Ref.[15]), where we can see a 
good agreement between both types of approaches. 

In Figure 2 are diplayed the results of the calculations of the electrons' 
mobility (left ordinate) in the steady state of doped ($n \simeq 5 \times 
10^{15} cm^{- 3}$) GaAs, as well as the momentum relaxation time (right 
ordinate), in the Ohmic regime, as a function of the reservoir temperature: 
It can be noticed a good agreement with the experimental data taken from 
three sources, namely, Refs.[16], [17] and [18]. 

Going over the particular case of the large-gap strongly polar GaN (in the 
cubic, i.e. zincblende, phase with $n \simeq 10^{18} cm^{- 3}$, and 
$T_{0} = 300K$), we can see in Figure 3 the increase of the electron-drift 
velocity with the electric field intensity, and a good agreement can be 
noticed in a comparison with the NMD-Monte Carlo simulation, the dots, taken 
from Ref.[19].

Finally, in Figure 4 we can accompany the evolution in time of the 
electron-drift velocity in doped ($n \simeq 10^{17} cm^{- 3}$) cubic GaN, 
where in the horizontal axis we do have the travelled distance at any time 
$t$, i.e. $v \, t$, and in the presence of an electric field intensity of 
$30 KV/cm$. We can notice, first, that the transient time is roughly of the 
order of $300$ femtoseconds and it can be observed the presence of a 
so-called ''velocity overshoot'' at roughly $100$ femtoseconds. The 
agreement with the NDM-Monte Carlo simulation, of Ref.[20], is very good. 
The calculations and figures are from Ref.[21].
\section{Concluding Remarks}
As noticed in the Introduction, in the previous article in this issue it was 
described the construction of a generalized higher-order-nonlinear 
hydrodynamics 
based on a nonequilibrium ensemble formalism. This means a mechano-statistical 
foundation in terms of a generalized nonequilibrium grand-canonical ensemble, 
in that way establishing a kind of unification of the microscopic dynamics and 
a mesoscopic hydrodynamics. In this follow up article we present some 
illustrations on the working of the formalism. In Section II we have considered 
the case of a system composed of two ideal fluids in interaction between them. 
It is used a truncated hydrodynamic description in which are included solely 
the density of energy, the density of particles, and the first flux (current) 
of particles. As discussed elsewhere this implies on restrictions on the 
characteristics of the hydrodynamic motion, basically the case of those 
restricted to long to intermediate wavelenghs \cite{a6,a22}, i.e. a 
hydrodynamics of first order. The continuity equations are derived 
[cf. Eqs.(22) to (24)] and to close them, in this truncated description, one 
needs to express the second-order flux in Eq.(24) in terms of the basic 
variables. 
In doing this we find a first source of nonlinearity as can be seen in Eq.(27). 
Moreover, the collision integrals on the right of these equations of evolution 
are highly nonlinear as shown, for the case of the one in the equation of 
evolution for the first flux, in Appendix A in the model of two ideal 
fluids in interaction, and in Appendix B for the case of a fluid of conduction 
electrons in doped semiconductors. In the first case the collision integral 
is composed of several contributions: One dependent on the gradient of the 
density, as it should appear in a mesoscopic hydrodynamics of this type 
\cite{a23}. Other accounting for nonlocal effects, and a local in space third 
one 
which is of the type of Maxwell's contribution \cite{a4,a7,a8}; cf. Eq.(A6). 
It is of a tensorial (rank 2) character and highly nonlinear in the 
nonequilibrium thermodynamic variables, and then dependent on the position 
and time.

Section III deals with the motion under flow (advective motion) for the case 
of the description of Section II. The equations of evolution for the density of 
particles and for the field of velocity - which are coupled - are obtained. 
In that way are derived, respectively, a generalized Maxwell-Cattaneo-type 
equation and a generalized Navier-Stokes-like equation. The nonlinearities 
are present, and the generalized diffusion coefficient is dependent on the 
local and instantaneous nonequilibrium thermodynamic state of the system.

Finally, in Section IV it is considered quantum transport in doped polar 
semiconductors, that is we deal with the homogenous current of the mobile 
electrons. In this case it is obtained in the steady state - that sets in 
after a very short (nanosecond scale) transient - when under the action of 
a constant electric field (that is, the space dependence carried on in the 
previous Sections is not present). It is derived the dependence of the 
nonequilibrium electron temperature (carrier's quasitemperature) on the 
electric field intensity as well as of the drift velocity (the current is 
proportional to it) in several cases: Comparison with computational 
modeling Monte Carlo calculations and with experimental data is done 
following a very good agreement. 
\begin{acknowledgments}
We acknowledge financial support from S\~ao Paulo State Research 
Foundation (FAPESP). ARV and RL are Brazil National Research Council 
(CNPq) research fellows.
\end{acknowledgments}
\appendix
\section{The Collision Integral}
The collision integral present on the right-hand side of Eq.(17), resulting 
from the collisions of the two types of particles via the potential 
${\mathcal V}(\left| {\mathbf r}_{j} - {\mathbf R}_{\mu} \right|)$ 
[cf. Eq.(1)] is given in the Markovian approximation \cite{a11,a24,a25,a26} 
by the expression 
\begin{eqnarray}
{\mathbf J}_{n}({\mathbf r},t) \simeq \int_{-\infty}^{0} dt' \, 
e^{\epsilon (t'-t)} \, \int d\Gamma \, \left\{ 
\hat{H}^{\prime}(\Gamma|t'-t)_{0}, 
\left\{ \hat{H}^{\prime}(\Gamma), \hat{{\mathbf I}}_{n}(\Gamma|{\mathbf r}) 
\right\} \right\} \, \bar{\rho}(t,0) \times \rho_{R} \, ,
\end{eqnarray}
where $\hat{H}^{\prime}(t'-t)_{0}$ indicates evolution under the dynamics 
generator by $\hat{H}_{0}$ - i.e. in the interaction representation -, and 
we use for the statistical distribution of the reservoir, $\rho_{R}$, a 
canonical 
one with temperature $\beta_{R}$. It can be noticed that the approximate 
(Markovian) scattering integral of Eq.(A1) is quadratic in the interaction 
strength and corresponds in the classical limit to the Golden Rule of 
Quantum Mechanics, and we recall that $\left\{ \hdots , \hdots \right\}$ stands 
for Poisson bracket. 

The lengthy but straightforward calculation of Eq.(1) provides several 
contributions to this scattering integral: one involving the gradient of 
the concentration (leading to a correction (renormalization) of the diffusion 
coefficient due to the collisions), other involving the gradient of the 
quasitemperature (implying in a cross-effect between thermal and material 
motion). Another one is related to introduce correlation effects in space 
of the collisional processes (i.e. nonlocal effects). Finally, the most 
relevant contribution is given by 
\begin{eqnarray}
{\mathbf J}_{n}({\mathbf r},t) \simeq - \frac{N_{R}}{m \, m_{r} \, V^{2}} 
\sqrt{\frac{\pi}{2}} \left( M \, \beta_{R} \right)^{3/2} 
\sum_{{\mathbf q}} \frac{|{\mathcal V}(q)|^{2}}{q} \, 
[ {\mathbf q} \, {\mathbf q} ] \, {\mathbf A}({\mathbf r},{\mathbf q}) \, ,
\end{eqnarray}
where
\begin{eqnarray}
{\mathbf A}({\mathbf r},{\mathbf q}) = n({\mathbf r},t)  
\left[ \frac{\beta({\mathbf r},t)}{2 \pi m} \right]^{3/2}
\int d^{3}p \, {\mathbf p} \exp \left\{ - 
\frac{\beta({\mathbf r},t)}{2 \, m} 
\left[ {\mathbf p} - m {\mathbf v}({\mathbf r},t) \right]^{2} - 
\frac{M \beta_{R}}{2 m^{2} q^{2}} 
\Big( {\mathbf q} \cdot {\mathbf p} \Big)^{2} \right\} \, ,
\end{eqnarray}
In Eqs.(A2) and (A3), ${\mathcal V}({\mathbf q})$ is the Fourier transform 
of the interaction potential between the two types of particles, $m_{r}$ is 
the reduced mass, $m_{r}^{- 1} = m^{- 1} + M^{- 1}$, and $N_{R}$ is the number 
of particles in the thermal bath.

Introducing a shift in coordinate ${\mathbf p}$, namely 
\begin{eqnarray}
{\mathbf p} = {\mathbf P} + {\mathbf b} \, ,
\end{eqnarray}
with
\begin{eqnarray}
\left[ \beta({\mathbf r},t) 1^{[2]} + \frac{M \beta_{R}}{m q^{2}} \, 
[ {\mathbf q} \, {\mathbf q} ] \right] \, {\mathbf b}({\mathbf r},t) = 
\beta \, m \, {\mathbf v}({\mathbf r},t) \, , 
\end{eqnarray}
and performing the calculations we arrive at the result that
\begin{eqnarray}
{\mathbf J}_{n}({\mathbf r},t) \simeq \left[ \Theta_{n}^{[2]}({\mathbf r},t) 
\right]^{- 1} \, {\mathbf I}_{n}({\mathbf r},t) \, ,
\end{eqnarray}
where $\Theta$ is a rank-two tensor with dimensions of time, playing the 
role of a tensorial Maxwell-relaxation time \cite{a4,a7,a8}, with its 
inverse given by 
\begin{eqnarray}
\left[ \Theta_{n}^{[2]}({\mathbf r},t) \right]^{- 1} &=& 
- \frac{n_{R} \beta_{R}}{m} \frac{m + M}{m}  
\left( \frac{\pi M \beta_{R}}{2} \right)^{1/2}  
\left[ \frac{m \beta({\mathbf r},t)}{
m \beta({\mathbf r},t) + M \beta_{R}} \right]^{3/2}  
K^{[2]}({\mathbf r},t) \, , 
\end{eqnarray}
where
\begin{eqnarray}
K^{[2]}({\mathbf r},t) = \frac{1}{V} \sum_{{\mathbf q}} 
\left| {\mathcal V}(q) \right|^{2} \, 
\frac{[{\mathbf q} \, {\mathbf q}]}{q} \, \exp \left\{ - 
\frac{m \, M \, \beta_{R} \, \beta({\mathbf r},t)}{
2 \, [ m \, \beta({\mathbf r},t) + M \, \beta_{R}]} \, 
\frac{({\mathbf q} \cdot {\mathbf v}({\mathbf r},t) )^{2}}{q^{2}} 
\right\} \, ,
\end{eqnarray}
with $n_{R}$ being the density of particles in the thermal bath. 

We stress that this tensorial Maxwell-relaxation time depends on position and 
time through its dependence on the basic nonequilibrium thermodynamic variables 
$\beta({\mathbf r},t)$ and ${\mathbf v}({\mathbf r},t)$: recalling that 
${\mathbf I}_{n}({\mathbf r},t) = n({\mathbf r},t) \, 
{\mathbf v}({\mathbf r},t)$, this time-relaxation-type contribution is not 
a linear one, but a highly nonlinear in ${\mathbf v}({\mathbf r},t)$. 
In fact, performing the integration in Eq.(A8) we find that 
$K_{ij} = 0$ for $i \neq j$, and there survive the diagonal terms 
\begin{eqnarray}
K_{xx}({\mathbf r},t) = K_{yy}({\mathbf r},t) = 
\xi(0) \, \Phi \left( \frac{m}{2} \tilde{\beta}({\mathbf r},t) \, 
v^{2}({\mathbf r},t) \right) - \frac{1}{2} K_{zz}({\mathbf r},t) \, , 
\end{eqnarray}
\begin{eqnarray}
K_{zz}({\mathbf r},t) = \xi(0) \, \left[ \frac{m}{2} 
\tilde{\beta}({\mathbf r},t) \, 
v^{2}({\mathbf r},t) \right]^{- 3/2} \, 
\gamma \left( \frac{3}{2}, \frac{m}{2} \tilde{\beta}({\mathbf r},t) \, 
v^{2}({\mathbf r},t) \right) \, , 
\end{eqnarray}
where
\begin{eqnarray}
\tilde{\beta}({\mathbf r},t) = \frac{M \, \beta_{R} \, \beta({\mathbf r},t)}{
m \, \beta({\mathbf r},t) + M \, \beta_{R}} \, , 
\end{eqnarray}
\begin{eqnarray}
\Phi \left( \frac{m}{2} \tilde{\beta}({\mathbf r},t) \, 
v^{2}({\mathbf r},t) \right) = 
\sum_{k=1}^{\infty} (- 1)^{k+1} \, \frac{2}{(2k - 1)(k - 1)!} \, 
\left( \frac{m}{2} \tilde{\beta}({\mathbf r},t) \, 
v^{2}({\mathbf r},t) \right)^{2(k-1)} \, ,  
\end{eqnarray}
\begin{eqnarray}
\xi(0) = \frac{1}{8 \, \pi^{2}} \int_{0}^{q_{0}} dq \, q^{3} \, 
\left| {\mathcal V}({\mathbf q}) \right|^{2} \, ,
\end{eqnarray}
with $q_{0}$ being a cut-off limit of the order of the inverse of the mean 
distance between particles of the system and the bath, $\gamma$ is the 
incomplete Gamma function \cite{a12}, and we can also alternatively write 
\begin{eqnarray}
K_{zz}({\mathbf r},t) = \xi(0) \sum_{0}^{\infty} (- 1)^{k} \, 
\frac{2}{\left( k + \frac{3}{2} \right) \, k!} 
\left( \frac{m}{2} \tilde{\beta}({\mathbf r},t) \, 
v^{2}({\mathbf r},t) \right)^{k} \, .
\end{eqnarray}
It may be noticed that we can write 
\begin{eqnarray}
\left[ \Theta_{n}^{[2]}({\mathbf r},t) \right]^{- 1} = 
\frac{1^{[2]}}{\Theta_{n}({\mathbf r},t)} + 
\left[ \overset{\circ}{\Theta}_{n}^{[2]}({\mathbf r},t) \right]^{- 1} ,
\end{eqnarray}
where
\begin{eqnarray}
\frac{1}{\Theta_{n}({\mathbf r},t)} = \frac{1}{3} Tr \left\{ 
\left[ \Theta_{n}^{[2]}({\mathbf r},t) \right]^{- 1} \right\} , 
\end{eqnarray}
and the last term is then traceless, and accounting for anisotropic effects. 
Furthermore, using Eq.(A10) there follows that
\begin{eqnarray}
\frac{1}{\Theta_{n}({\mathbf r},t)} = \frac{1}{3} 
\Big( K_{xx}({\mathbf r},t) + K_{yy}({\mathbf r},t) + 
K_{zz}({\mathbf r},t) \Big) = 
\frac{2}{3} \xi(0) \, \Phi \left( \frac{m}{2} \tilde{\beta}({\mathbf r},t) \, 
v^{2}({\mathbf r},t) \right) . 
\end{eqnarray}
Conserving only this contribution in Eq.(A14), linearizing in 
${\mathbf v}({\mathbf r},t)$ all expressions, say, the case of a low 
density flux, assuming a good thermal contact between system and 
reservoir, such that $\beta({\mathbf r},t) \simeq \beta_{R}$, and small 
amplitude movement, namely $n({\mathbf r},t) \simeq n$, we are left with a 
typical Maxwell-type contribution 
\begin{eqnarray}
{\mathbf J}_{n}({\mathbf r},t) \simeq 
- {\mathbf I}_{n}({\mathbf r},t) / \Theta_{n} ,
\end{eqnarray}
where 
\begin{eqnarray}
\frac{1}{\Theta_{n}} = \sqrt{\frac{\pi}{2}} n_{R} \, \beta_{R}^{3/2} \, 
\sqrt{\frac{M}{m (m + M)}} \, \xi(0) .
\end{eqnarray}
\section{Energy and Momentum Relaxation, Eqs.(47) and (48)}
The collision integrals are in this case given by 
\begin{eqnarray}
J_{j}^{(2)}(t) = \left( \frac{1}{{\textrm i} \, \hbar} \right)^{2} 
\int_{\- \infty}^{t} dt' \, e^{\epsilon (t'-t)} \, Tr \left\{ 
\left[ \hat{H}^{\prime}(t')_{0} , \left[ \hat{H}^{\prime}, \hat{A}_{j} 
\right] \right] \, \bar{\rho}(t,0) \right\} ,
\end{eqnarray}
where $\hat{A}_{1} \equiv \hat{H}_{e}$ for energy and 
$\hat{A}_{2} \equiv {\mathbf P}$ for momentum, and $\hat{H}^{\prime}$ is 
Fr\"ohlich electron-phonon interaction; subindex nought indicates time 
dependence in interaction representation. These operators are 
\begin{eqnarray}
\hat{H}_{e} = \sum_{{\mathbf k}} \frac{\hbar^{2} \, k^{2}}{2 \, m^{\ast}} 
c_{_{{\mathbf k}}}^{\dagger} \, c_{_{{\mathbf k}}} , 
\end{eqnarray}
\begin{eqnarray}
\hat{H}^{\prime} = \sum_{{\mathbf k}{\mathbf q}} {\mathcal C}_{{\mathbf q}} \, 
c_{_{{\mathbf k}+{\mathbf q}}}^{\dagger} \, c_{_{{\mathbf k}}} \, \left( 
b_{_{{\mathbf q}}} + b_{_{- {\mathbf q}}}^{\dagger} \right) , 
\end{eqnarray}
\begin{eqnarray}
{\mathbf P} = \sum_{{\mathbf k}} \hbar \, {\mathbf k} \,  
c_{_{{\mathbf k}}}^{\dagger} \, c_{_{{\mathbf k}}} , 
\end{eqnarray}
where $c_{_{{\mathbf k}}}$ ($c_{_{{\mathbf k}}}^{\dagger}$) are 
annihilation (creation) operators of electrons in band state 
${\mathbf k}$, and the effective mass approximation is taken; 
$b_{_{{\mathbf q}}}$ ($b_{_{{\mathbf - q}}}^{\dagger}$) are annihilation 
(creation) operator of LO phonons in mode ${\mathbf q}$, and the matrix element 
in Fr\"ohlich interaction is 
\begin{eqnarray}
{\mathcal C}_{{\mathbf q}} = - {\textrm i} \frac{\alpha^{1/2}}{q} , 
\end{eqnarray}
$\alpha$ being Fr\"ohlich coupling constant.

After some calculation one arrives at the results that 
\begin{eqnarray}
J_{E}^{(2)}(t) &=& \frac{2 \, \pi}{\hbar} \sum_{{\mathbf k}{\mathbf q}} 
\left| {\mathcal C}_{{\mathbf q}} \right|^{2} \, 
\left( \varepsilon_{_{{\mathbf k}+{\mathbf q}}} - \varepsilon_{_{{\mathbf k}}} 
\right) \Bigg\{ \Big( \nu_{_{{\mathbf q}}}(t) + 1 \Big) \, 
f_{_{{\mathbf k}}}(t) \, 
\Big( 1 - f_{_{{\mathbf k}+{\mathbf q}}}(t) \Big) \, 
\delta \Big( \varepsilon_{_{{\mathbf k}+{\mathbf q}}} - 
\varepsilon_{_{{\mathbf k}}} 
+ \hbar \, \omega_{_{{\mathbf q}}} \Big) \nonumber \\
&&- \,
\nu_{_{{\mathbf q}}}(t) \, f_{_{{\mathbf k}}}(t) \, 
\Big( 1 - f_{_{{\mathbf k}+{\mathbf q}}}(t) \Big) \, 
\delta \Big( \varepsilon_{_{{\mathbf k}+{\mathbf q}}} - 
\varepsilon_{_{{\mathbf k}}} 
- \hbar \, \omega_{_{{\mathbf q}}} \Big) \Bigg\} ,
\end{eqnarray}
\begin{eqnarray}
J_{{\mathbf P}}^{(2)}(t) &=& \frac{2 \, \pi}{\hbar} 
\sum_{{\mathbf k}{\mathbf q}} 
\left| {\mathcal C}_{{\mathbf q}} \right|^{2} \, \hbar \, {\mathbf q} \, \, 
\Bigg\{ \Big( \nu_{_{{\mathbf q}}}(t) + 1 \Big) \, f_{_{{\mathbf k}}}(t) \, 
\Big( 1 - f_{_{{\mathbf k}+{\mathbf q}}}(t) \Big) \, 
\delta \Big( \varepsilon_{_{{\mathbf k}+{\mathbf q}}} - 
\varepsilon_{_{{\mathbf k}}} 
+ \hbar \, \omega_{_{{\mathbf q}}} \Big) \nonumber \\
&&- \,
\nu_{_{{\mathbf q}}}(t) \, f_{_{{\mathbf k}}}(t) \, 
\Big( 1 - f_{_{{\mathbf k}+{\mathbf q}}}(t) \Big) \, 
\delta \Big( \varepsilon_{_{{\mathbf k}+{\mathbf q}}} - 
\varepsilon_{_{{\mathbf k}}} 
- \hbar \, \omega_{_{{\mathbf q}}} \Big) \Bigg\} ,
\end{eqnarray}
where, in a nondegenerate-like limit  acceptable in the usual experimental 
conditions 
\begin{eqnarray}
f_{_{{\mathbf k}}}(t) = A(t) \, \exp \left\{ 
- \beta(t) \frac{\hbar^{2}}{2 \, m^{\ast}} \left( {\mathbf k} - 
{\mathbf k}_{_{D}} \right)^{2} \right\} ,
\end{eqnarray}
a shifted Maxwell-Boltzmann-like instantaneous distribution, with 
$\hbar \, {\mathbf k}_{_{D}} = m \, {\mathbf v}$, 
${\mathbf P} = n \, {\mathbf v}$, $A(t)$ ensures the normalization to $N$, 
$\omega_{0}$ is the LO phonons dispersionless frequency (Einstein model), 
and [cf. Eq.(54)]
\begin{eqnarray}
\nu_{_{{\mathbf q}}}(t) = \Big[ \exp \left\{ 
\beta_{LO}(t) \, \hbar \, \omega_{o} \right\} - 1 \Big]^{- 1} .
\end{eqnarray}
It can be noticed that Eqs.(B6) and (B7) are of the form of the Golden Rule of 
Quantum Mechanics averaged over the noequilibrium ensemble. Performing the 
integrations one arrives to the results presented in Eqs.(47) and (48).
%
%
%

%
%
%
%
%
%
%
\section*{Figure Captions}
%
%
\begin{figure}[htbp]
\includegraphics[width=8.5cm]{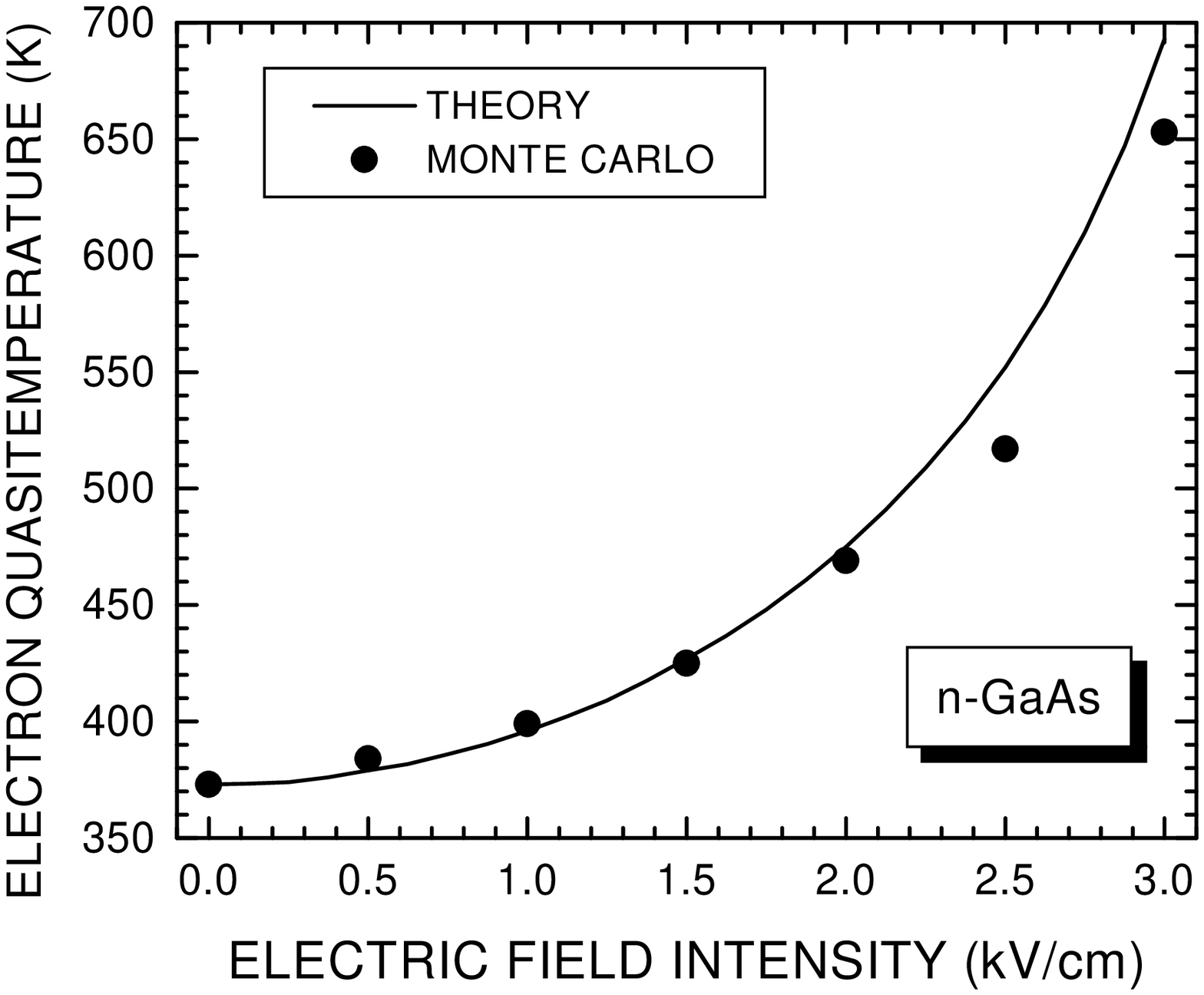}
\caption{The electron quasitemperature $vs.$ the electric field intensity, 
in the steady state of $n$-GaAs, comparing the results of the NESEF-based 
kinetic theory with a Monte Carlo 
simulation (full circles from ref.[15]).}
\end{figure}
%
%
\begin{figure}[htbp]
\includegraphics[width=11.5cm]{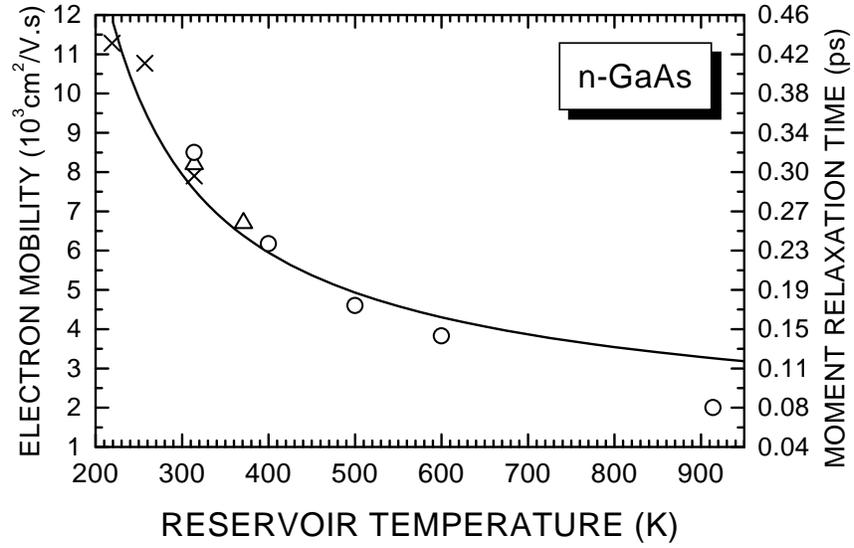}
\caption{Electron mobility (and momentum relaxation time) in $n$-GaAs for 
different values of the reservoir temperature, comparing the results 
of the NESEF-based kinetic theory with experimental data (up triangle 
from Ref.[14], $\times$ from Ref.[17], empty circles from Ref.[18]).}
\end{figure}
%
%
\begin{figure}[htbp]
\includegraphics[width=11.5cm]{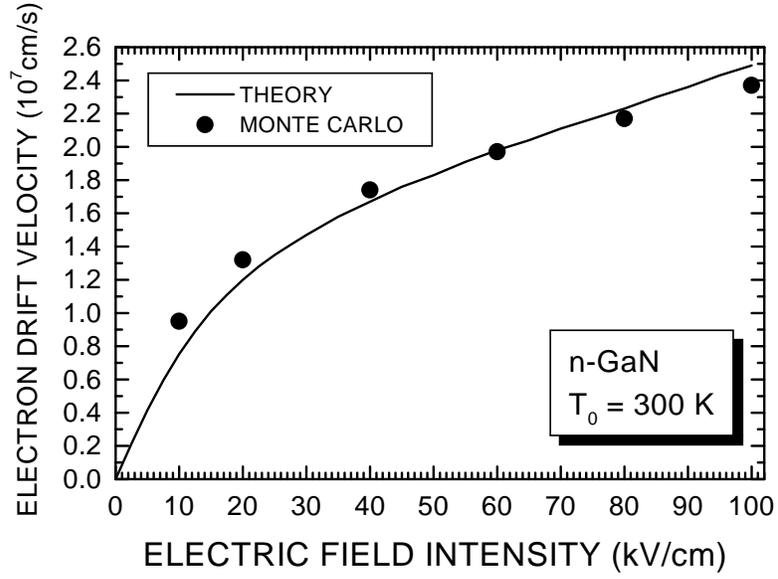}
\caption{Electron-drift velocity $vs.$ electric field intensity in $n$-GaN, 
comparing the results of the NESEF-based kinetic theory with 
a Monte Carlo simulation (Full circle from Ref.[19]), with $T_{0} = 300K$.}
\end{figure}
%
%
\begin{figure}[htbp]
\includegraphics[width=11.5cm]{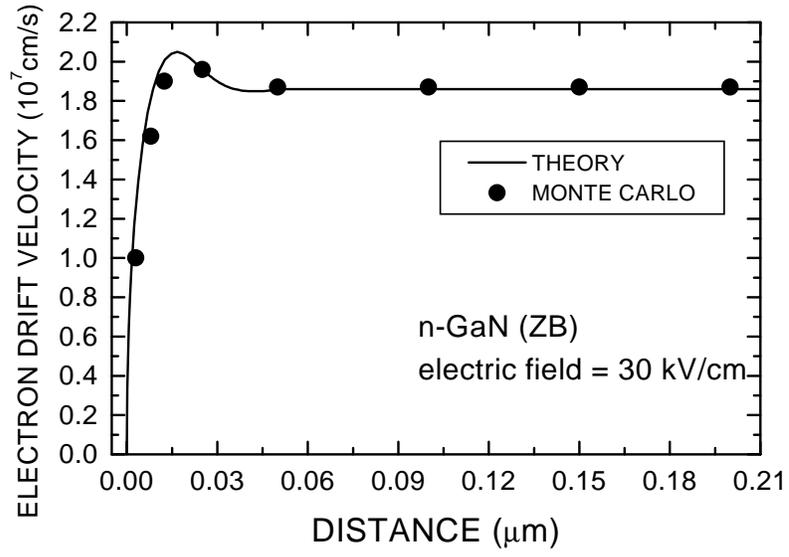}
\caption{Evolution of the electron-drift velocity in $n$-GaN in terms of 
the travelled distance, comparing the results of the NESEF-based kinetic 
theory with a Monte Carlo simulation (Full circles from Ref.[20]): 
zincblende $n$-GaN for an electric field intensity 30kV/cm and $T_{0} = 300K$.}
\end{figure}
\end{document}